\documentclass{article}

\usepackage{microtype}
\usepackage{graphicx}
\usepackage{subfigure}
\usepackage{booktabs} 
\usepackage{multirow}
\usepackage{float}
\usepackage{placeins}
\usepackage[version=4]{mhchem}  
\usepackage[numbers,sort&compress,super]{natbib}

\setcitestyle{numbers,sort&compress,super}

\usepackage{hyperref}

\usepackage[accepted]{icml2024}

\usepackage{amsmath}
\usepackage{amssymb}
\usepackage{mathtools}
\usepackage{amsthm}

\usepackage[capitalize,noabbrev]{cleveref}
\usepackage{fancyhdr}
\usepackage{multicol}

\theoremstyle{plain}

\theoremstyle{definition}

\theoremstyle{remark}

\usepackage[textsize=tiny]{todonotes}

\icmltitlerunning{Uni-ELF: A Multi-Level Representation Learning Framework for Electrolyte Formulation Design}

\begin{document}

\twocolumn[
\icmltitle{Uni-ELF: A Multi-Level Representation Learning Framework for Electrolyte Formulation Design}




\begin{icmlauthorlist}
\icmlauthor{Boshen Zeng}{dp,pku}
\icmlauthor{Sian Chen}{dp}
\icmlauthor{Xinxin Liu}{dp}
\icmlauthor{Changhong Chen}{dp}
\icmlauthor{Bin Deng}{dp}
\icmlauthor{Xiaoxu Wang}{dp}
\icmlauthor{Zhifeng Gao}{dp}
\icmlauthor{Yuzhi Zhang}{dp,aisi}
\icmlauthor{Weinan E}{aisi,pkuml}
\icmlauthor{Linfeng Zhang}{dp,aisi}
\end{icmlauthorlist}

\icmlaffiliation{dp}{DP Technology, Beijing 100080, P. R. China}
\icmlaffiliation{aisi}{AI for Science Institute, Beijing 100080, P. R. China}
\icmlaffiliation{pku}{Peking University, Beijing 100871, P. R. China}
\icmlaffiliation{pkuml}{Center for Machine Learning Research and School of Mathematical Sciences, Peking University, Beijing, China}

\icmlcorrespondingauthor{Linfeng Zhang}{linfeng.zhang.zlf@gmail.com}


\icmlkeywords{Machine Learning}

\begin{onecolumn}
\begin{center}
\begin{abstract}
\vskip 0.3in
Advancements in lithium battery technology heavily rely on the design and engineering of electrolytes. However, current schemes for molecular design and recipe optimization of electrolytes lack an effective computational-experimental closed loop and often fall short in accurately predicting diverse electrolyte formulation properties. In this work, we introduce Uni-ELF, a novel multi-level representation learning framework to advance electrolyte design. Our approach involves two-stage pretraining: reconstructing three-dimensional molecular structures at the molecular level using the Uni-Mol model, and predicting statistical structural properties (e.g., radial distribution functions) from molecular dynamics simulations at the mixture level. Through this comprehensive pretraining, Uni-ELF is able to capture intricate molecular and mixture-level information, which significantly enhances its predictive capability. As a result, Uni-ELF substantially outperforms state-of-the-art methods in predicting both molecular properties (e.g., melting point, boiling point, synthesizability) and formulation properties (e.g., conductivity, Coulombic efficiency). Moreover, Uni-ELF can be seamlessly integrated into an automatic experimental design workflow. We believe this innovative framework will pave the way for automated AI-based electrolyte design and engineering.

\looseness=-1
\end{abstract}
\end{center}
\end{onecolumn}

\twocolumn
]


\printAffiliationsAndNotice{}  


\section{Introduction}

Lithium-based rechargeable batteries are a cornerstone of modern energy storage technologies, offering exceptional potential for high energy density, rapid charging capabilities, and longevity. Functioning as an ionic conductor and electronic insulator between electrodes while maintaining stability under extreme chemical conditions, the electrolyte, which interfaces with every other component, plays a vital role in battery operation\cite{science-electrolyte-2022, nature2023-wangcs, nature2024-fanxl}. As we enter the era of high-energy-density batteries that place higher demands on electrolytes, especially with high-voltage cathode materials\cite{goodenough_challenges_2010, NatRC2024-wangcx} and high-energy-density anode materials like lithium metal\cite{cheng2017, NE2020-BaoZN-electrolyte-Limetal}, the design and engineering of electrolytes emerge as the main challenges. Current electrolyte systems based on ethylene carbonate (EC) are increasingly inadequate for these next-generation energy storage solutions\cite{xuk2014, chemsocrev2023-Chenjun}. Consequently, breakthroughs in materials and chemistries crucial for next-generation batteries hinge on mastering electrolyte design.

\begin{figure*}[h!]
\centering
\includegraphics[width=0.9\textwidth]{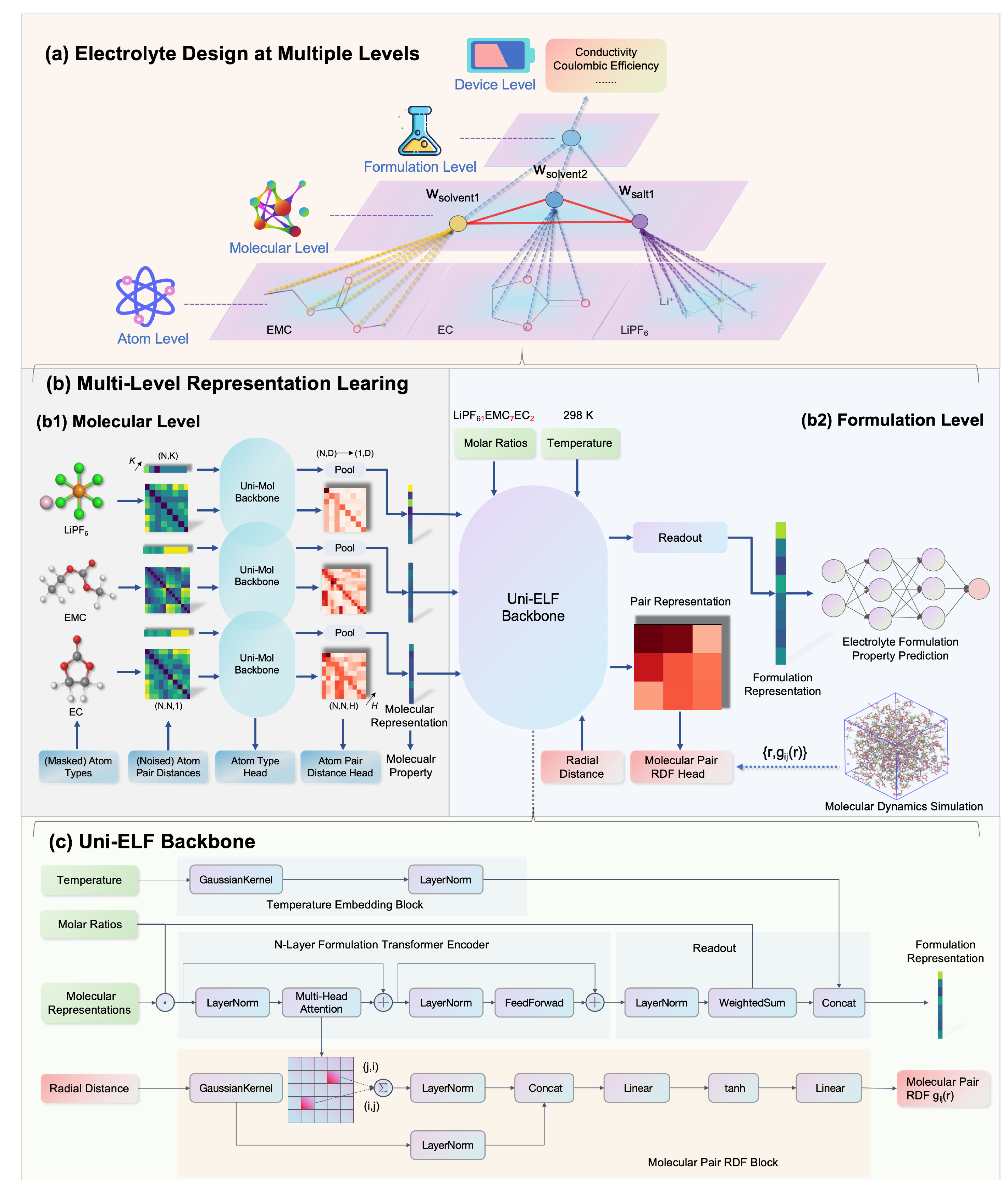}
\caption{\textbf{Electrolyte formulation representation learning framework.}
\textbf{a, Electrolyte design at multiple levels.} At the atomic level, individual atoms and their interactions form molecular geometric structures, creating molecular-level representations. Based on these, individual molecular species, their proportions, and their interactions (depicted by red lines) within the mixtures create formulation-level representations, which are then used to predict device-level properties.
\textbf{b, Multi-level representation learning: b1.} Molecule-level representations are learned through self-supervised tasks, including recovering masked atom types and denoising atom pair distances. {\bf b2.} These refined representations are then fed with mixture ratios into the Uni-ELF backbone.
\textbf{c, Uni-ELF backbone model architecture.} The Uni-ELF model is based on a transformer encoder design. Molar ratios are used as weights for molecular representations, and pair representations are maintained for mixture-level pretraining. Symmetrical elements in the pair representation matrix are summed and combined with the radial features obtained from the Gaussian kernel. These combined features are then used to predict radial distribution functions (RDFs), a pretraining task to recover the structural properties of the mixed system.}
\label{fig:model}
\end{figure*}

The research and development of electrolytes present two primary challenges: innovating molecular design and manipulating electrolyte formulation. These challenges stem from the need to fine-tune the electrolyte's conductivity\cite{conductivity-LIPF6ECEMC-2001, conductivity-EES2012, formulation-2022}, solubility\cite{formulation-2024, solubility-DN, solubility-dielectric}, stability\cite{NE2020-BaoZN-electrolyte-Limetal, stability-JACS-ChenX}, and compatibility with electrode materials\cite{nature2023-wangcs, nature2024-fanxl} to meet stringent performance criteria. Unlike other fields, such as drug design, which mainly focus on the design and synthesis of monomeric small molecules, the design at the electrolyte formulation level is particularly crucial. This involves providing recommendations and predictions for the mixing ratio of molecules, including lithium salts, solvents, and functional additives. The interplay between these different components can significantly affect the energy density, cycle life, and overall performance of the batteries\cite{eScience-2023-rev, SmartMat-2023-rev}. The variety of molecular space further exacerbates the challenge for potential candidates and the abundance of mixing possibilities, especially in multi-component systems\cite{formulation-2022, formulation-2024, chem-rev-MD-2022}.
We refer to Figure~\ref{fig:model}(a) for an illustration of electrolyte design at multiple levels.

The methodologies that heavily rely on trial-and-error lack the efficiency required for the rapid development of electrolyte systems. Over the past few decades, progress in computational approaches such as density functional theory (DFT)\cite{DFT-01, DFT-02} and molecular dynamics\cite{MD-Tuckerman-book} has enabled the deciphering of dynamic behaviors at the electronic and atomic levels, thereby deducing macroscopic properties through statistical mechanics. However, the complex nature inside batteries, especially across multiple scales, hinders a complete understanding of mechanisms, the development of highly capable and predictive simulators, and the realization of an ultimately rational design scheme\cite{chem-rev-MD-2022}. Moreover, the computational costs originating from the curse of dimensionality—the $O(N^3)$ complexity of DFT with respect to atoms, and the need for adequate sampling of necessary microscopic states—are not capable of matching the high-throughput screening in industrial research and development scenarios. 

On the other hand, data-driven schemes such as quantitative structure-property relationships have been developed, wherein the molecular representation is attained through feature engineering\cite{mathieu_reliable_2016,mansouri_opera_2018,bouteloup_improved_2018, bouteloup_predicting_2019,bradford_chemistry-informed_2023,cuiyi_panas_2023,lee_emin_2024}. The manual design of features or descriptors requires extensive domain knowledge and tends to be disadvantageous when confronting large-scale and high-dimensional problems. Furthermore, the scarcity of informative data makes the transferability of data-driven models uncertain. The rapid growth of deep learning techniques, especially molecular representation learning along with the pretraining-finetuning paradigm, has alleviated this problem\cite{dmpnn,pagtn,gcn,gat}. Among these methods, the Uni-Mol framework\cite{zhou2023unimol}, which properly incorporates the 3D information of a molecule, has achieved widespread success in a series of chemistry and material science fields, including small organic molecules\cite{lu2023unimol+}, organic light-emitting diodes\cite{cheng2023automatic}, and metal-organic frameworks\cite{wang_comprehensive_2024}, mostly focusing on the relationship between individual molecules and their properties. However, a similar approach has been lacking at the level of formulations, for which existing attempts are primarily based on traditional regression methodologies and conventional machine learning models such as random forest\cite{cuiyi_panas_2023} and XGBoost\cite{Chen_2016}.

In this study, we introduce the Universal Electrolyte Formulation (Uni-ELF) framework, which excels in predicting electrolyte properties and designing electrolyte formulations through a multi-level pretraining scheme: at the molecular level, it reconstructs three-dimensional molecular structures using the Uni-Mol model; while at the mixture level, it predicts statistical structural properties, such as radial distribution functions, derived from molecular dynamics simulations. 
Systematic experiments demonstrate that, after pretraining, Uni-ELF exceeds existing state-of-the-art (SOTA) methods across a broad spectrum of tasks, accurately predicting crucial properties at both molecular and mixture levels. The performance of Uni-ELF is anticipated to further improve by integrating physics-driven modeling and leveraging high-quality data acquired through autonomous experiments. We posit that Uni-ELF not only represents an innovative approach to unifying representation learning tasks for electrolytes across different levels but also serves as a timely and effective tool for intelligent battery design at the industrial scale.

\section{Multi-Level Representation Learning}

\begin{figure*}[ht!]
\centering
\includegraphics[width=\textwidth]{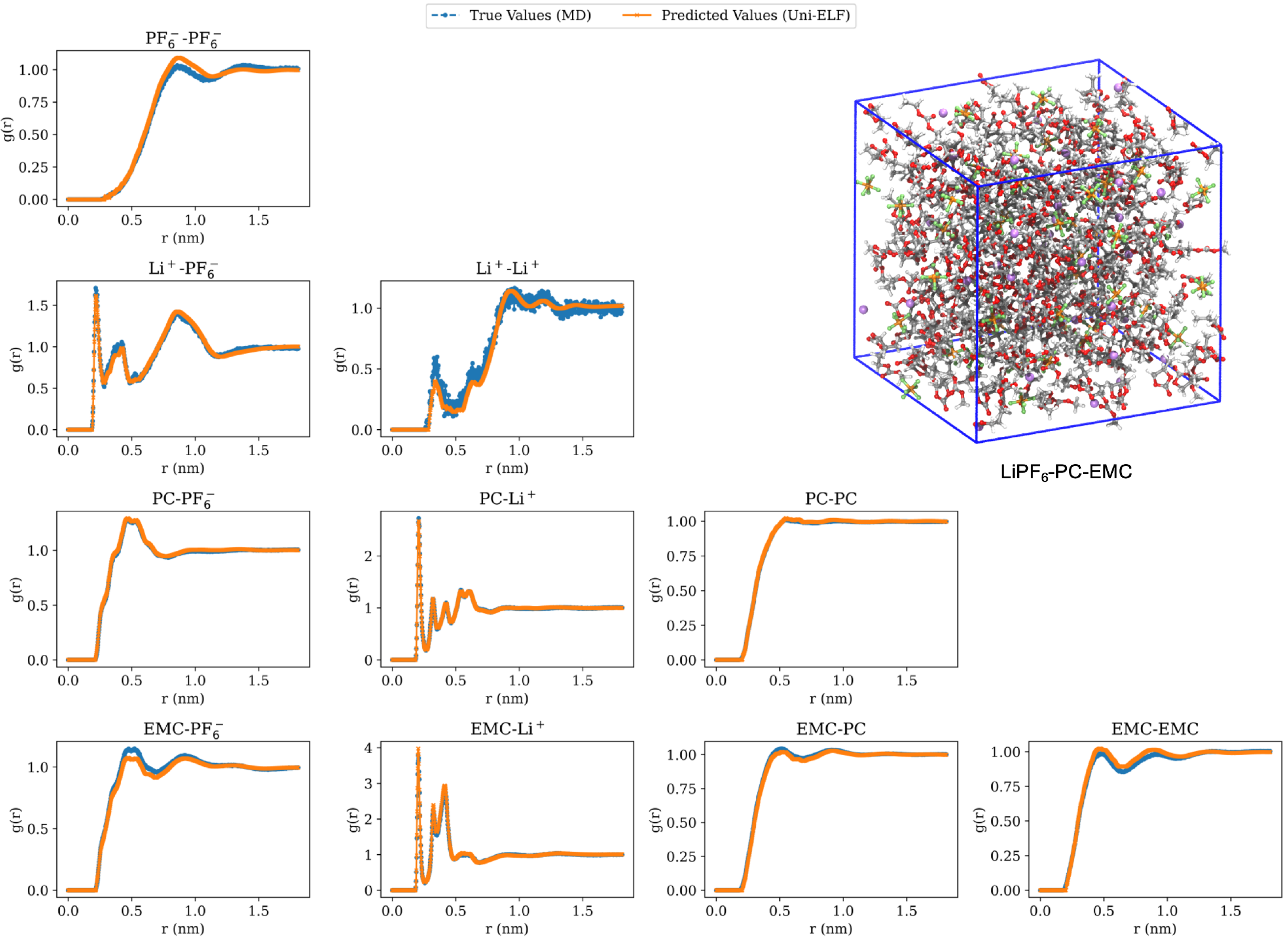}
\caption{\textbf{Prediction of molecular pairwise RDFs as a formulation-level pretraining task, using the LiPF$_6$/PC/EMC system with a molar ratio of n(Li$^+$) : n(PF$_6^-$) : n(PC) : n(EMC) = 0.12 : 0.12 : 0.54 : 0.22 as an example.} The plots compare the true values obtained from molecular dynamics (MD) simulations (blue) with the predicted values from the Uni-ELF model (orange) for various molecular pairs: PF$_6^-$, Li$^+$, PC, and EMC, including all pairwise combinations forming a lower triangular matrix. The right panel illustrates the system configuration. The strong agreement between predicted and true RDFs demonstrates the accuracy of the Uni-ELF model during pretraining.}
\label{fig:rdf}
\end{figure*}

In the Uni-ELF scheme, we first acquire molecular and formulation representations through pretraining. After this phase, the model can be fine-tuned for various tasks by linking these representations to different fitting networks. For a visual overview, please refer to Figure~\ref{fig:model}(b), which illustrates representation learning at both the molecular and formulation levels.

\subsection{Representation Learning at Molecular Level} 
The molecule-level representation learning approach is built upon Uni-Mol\cite{zhou2023unimol}, a three-dimensional molecular representation framework that leverages self-supervised pretraining to reconstruct molecular structures. As illustrated in Figure~\ref{fig:model}(b1), molecules, including key electrolyte components such as lithium hexafluorophosphate (LiPF$_6$), ethylmethyl carbonate (EMC), ethylene carbonate (EC), and propylene carbonate (PC), are encoded using their three-dimensional coordinates and atom types. These encodings are refined to generate atom-pair representations and atomic representations. During pretraining, the model unmaskes atom types and denoises atom pair distances. Following pretraining on 209 million molecular conformations, we employ average pooling over all atomic representations to derive molecular representations, which are subsequently used for predicting molecular properties or serving as input for the formulation-level model.

In greater detail, the 3D structures of input molecules are generated using the MMFF94 force field\cite{MMFF_Halgren1996} from RDKit \cite{RDkit_Landrum2013}. The Uni-Mol framework \cite{zhou2023unimol} serves as the encoder, comprising 15 layers with an embedding dimension of 512 and a feedforward network dimension of 2048. Each encoder layer is equipped with 64 attention heads, utilizing GELU\cite{hendrycks2016gaussian} for activation and tanh\cite{lecun2015deep} for pooler activation. The [CLS] token \cite{devlin-etal-2019-bert}, a virtual atom positioned at the center of mass of the molecule, is preserved to represent the entire structure in Uni-Mol. This design enables the model to capture long-range interactions between atoms, particularly within larger molecules. In tasks involving molecular charge distribution—such as predicting the dielectric constant and refractive index—atomic representations are used to simultaneously predict Gasteiger charges \cite{gasteiger1980iterative} within the molecule, thereby enhancing the model's ability to capture relevant electrostatic properties. The mean squared error (MSE) of the predicted charges is included as a loss term, with a weight of 0.1.

\subsection{Representation Learning at Formulation Level}
To enhance predictive capabilities, the formulation model should integrate specific inductive biases. Recognizing that entities are characterized not only by their intrinsic properties but also by their interactions with other entities, the model must distinguish between identical molecular species in varying contexts. Additionally, it should uphold permutation invariance for molecular input sequences, ensuring consistent output regardless of the order of inputs.

To achieve these goals, we designed the Uni-ELF backbone employing a transformer encoder architecture, as depicted in Figure~\ref{fig:model}(b2, c). At the formulation level, the model processes molecular representations weighted by their molar ratios, refining the representations of both individual molecular species and their interactions. These refined representations are then aggregated on the basis of their molar ratios. For tasks that involve environmental temperature, we introduce a temperature embedding block utilizing a Gaussian kernel. This block encodes temperature values through a set of evenly distributed Gaussian basis functions with specified means and standard deviations.

The model undergoes pre-training to predict solution structures, thereby learning formulation representations. Given the scarcity of experimental data, we supplement this with physical modeling to provide an additional source of structural data for transfer learning. Within the Uni-ELF framework, molecular dynamics simulations generate extensive data on the trajectories of solution particles. These trajectories are statistically averaged to extract the structural characteristics of the solution. 
Specifically, the radial distribution functions (RDFs) provide the density probability for a particle to have a neighbor at a given distance \( r \), revealing the fine structure of the solution. The RDFs of molecular pairs (detailed in Supplementary Information) are particularly suitable for edge-level tasks using pair representations in the transformer encoder, thus chosen as the data for the pretraining task.

During pretraining, Uni-ELF receives not only molecular species and their molar ratios but also a range of radial distance values \(r\). These radial distances are embedded using a Gaussian kernel. The model maintains pairwise representations of molecular species, leveraging the symmetry inherent in the RDF between molecules. Specifically, it sums the attention representations of matrix elements \(i,j\) and \(j,i\) to form the pairwise representation. This summed representation is then concatenated with the embedded radial distance values to predict the RDF \(g_{ij}(r)\) for the molecular pair \(i, j\) at a given radial distance \(r\).

In predicting RDFs, the model achieves a final test set root mean square error (RMSE) of 0.06. As illustrated in Figure \ref{fig:rdf}, the strong concordance between the predicted and true RDFs for a test set comprising the LiPF$_6$/PC/EMC system underscores the accuracy of the Uni-ELF model during pretraining. This high level of accuracy in reproducing the structural information of the formulations indicates a promising transfer of these learned representations to downstream property prediction tasks.

We refer to the Supplementary Information for more details of formulation-level model architecture, pretraining scheme, as well as molecular dynamics simulations.

\section{Results on Downstream Tasks}
\subsection{Molecule-Level Tasks}
\begin{figure}[h!]
\centering
\includegraphics[width=\columnwidth]{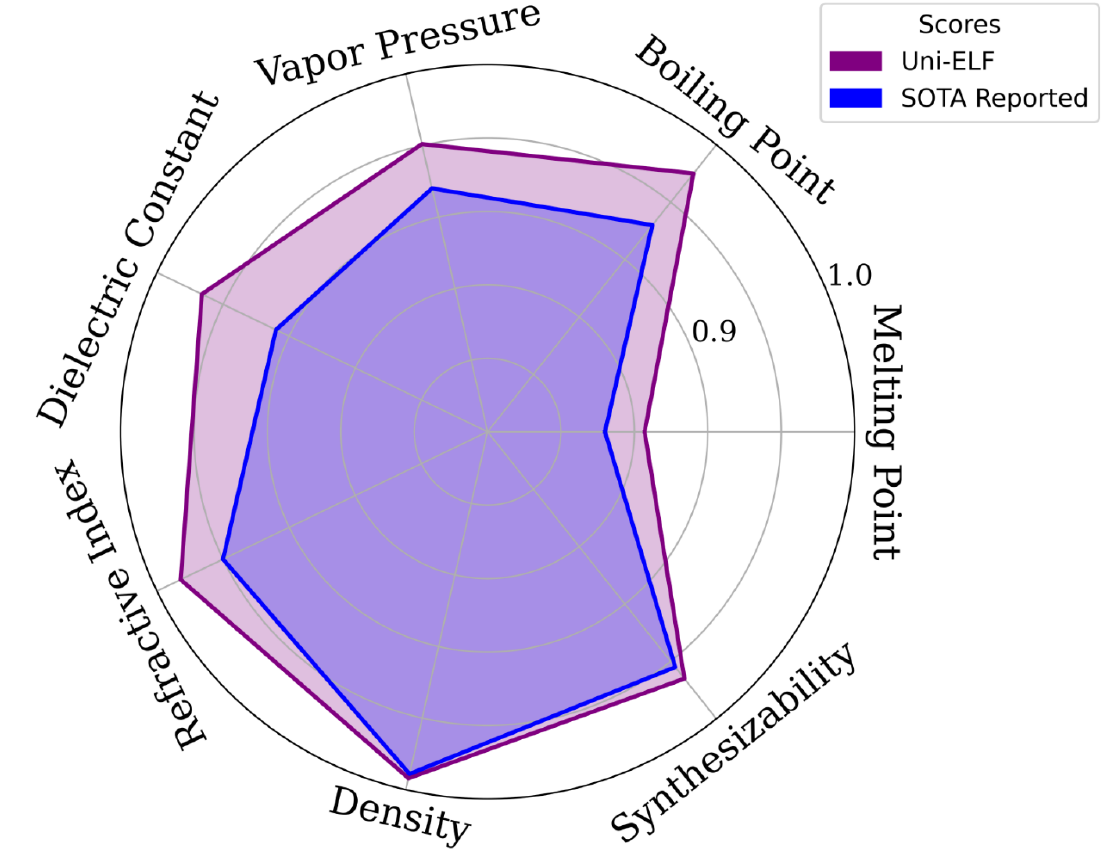}
\caption{\textbf{Comparative performance in predicting molecular properties for electrolyte design.} Uni-ELF (in purple) surpasses previously reported state-of-the-art (SOTA) methods (in blue) in predicting seven molecular properties (melting point, boiling point, vapor pressure, dielectric constant, refractive index, density on R² scores, and synthesizability on the AUC), which are essential for the inverse molecular design of electrolytes. Each concentric circle represents an interval of 0.05, with the outermost boundary corresponding to a perfect score of 1.0.}
\label{fig:molecule}
\end{figure}

We begin by leveraging the molecular representation capabilities of Uni-ELF to predict properties critical for electrolyte design. As illustrated in Figure \ref{fig:molecule}, Uni-ELF demonstrates superior performance compared to state-of-the-art methods. For melting point prediction, it achieves an R² of 0.857 and an RMSE of 34.31 °C, outperforming the previous benchmark of R² 0.830 and RMSE 36.88 °C\cite{mi_melting_2021}. In the prediction of boiling points and vapor pressures, Uni-ELF surpasses the OPERA model\cite{mansouri_opera_2018}, with an R² of 0.975 and an RMSE of 13.49 °C for boiling points, and an R² of 0.951 and an RMSE of 0.79 Log mm/Hg for vapor pressures. Additionally, it exceeds QSPR models in predicting dielectric constant, refractive index, and density, achieving R² values of 0.966, 0.982, and 0.992, with corresponding RMSEs of 2.70, 0.082, and 0.025 g/cm³, respectively\cite{bouteloup_predicting_2019,bouteloup_improved_2018,mathieu_reliable_2016}. These results underscore the advantage of representation learning over traditional QSPR methods in predicting molecular properties.

To further explore the model’s capability in identifying promising electrolyte molecules, we evaluate its performance on molecular synthesizability prediction. Predicting the synthesizability of new molecules is a challenging task, often dependent on the intuition and experience of chemists. Lee et al.\cite{lee_emin_2024} curated a dataset from QM9\cite{qm9}, comprising 126,405 entries, to assess molecular synthesizability. They classified QM9 molecules as synthesizable if they were listed in either the PubChem\cite{kim2023pubchem} or eMolecules\cite{emolecules} databases, while unlisted molecules were presumed unsynthesizable. In this task, our model achieves an area under the curve (AUC) of 0.965, surpassing the previous best AUC of 0.955\cite{lee_emin_2024}. Although the absence of a molecule in these databases does not definitively indicate unsynthesizability, it provides valuable insights into the relative ease or difficulty of synthesis. By coupling the conditions required for electrolytes, such as a wide liquid range and solubility for lithium salts, with trained models for melting point, boiling point, dielectric constant, and synthesizability, our approach offers a robust reference for evaluating the potential suitability and synthetic feasibility of virtually generated molecules as electrolytes.

We refer to the Supplementary Information for more details and additional benchmarks on molecular-level tasks, confirming the superior performance of Uni-ELF in various cases.

\begin{table*}[ht!]
    \centering
    \begin{tabular}{lcccc}
        \toprule
        \textbf{Method} & \textbf{Configuration} & \textbf{LCE} & \multicolumn{2}{c}{\textbf{Liquid Electrolyte Conductivity}} \\
        \cmidrule{4-5}
        & & & \textbf{Random Split} & \textbf{Group Split} \\
        \midrule
        Kim et al.\cite{cuiyi_panas_2023} & Random forest& 0.58 & & \\
        One-hot embedding & XGBoost & 0.246 (0.033) & 1.53 (0.15) & 3.15 (1.12) \\
        Morgan fingerprint & XGBoost & 0.231 (0.027) & 1.35 (0.08) & 3.11 (0.49) \\
        Uni-Mol fingerprint & XGBoost & 0.228 (0.027) & 1.23 (0.09) & 2.82 (0.56) \\
        \multirow{2}{*}{Uni-ELF} & w/o pretraining & 0.215 (0.021) & 0.53 (0.02) & 2.49 (0.63) \\
        & w/ pretraining & \textbf{0.184 (0.019)} & \textbf{0.50 (0.02)} & \textbf{2.15 (0.35)} \\
        \bottomrule
    \end{tabular}
    \caption{\textbf{RMSE results on the Coulombic efficiency and liquid electrolyte conductivity datasets for different methods and configurations, with the best RMSE denoted in bold.} The random split column represents the data randomly divided into training and test sets, while the group split column represents the data grouped by formulation systems containing identical sets of molecular species and randomly split into training and test sets according to their group. Results are reported as the mean of three independent experiments, with standard deviation in parentheses.}
    \label{tab:rmse_results}
\end{table*}

\begin{figure*}[h!]
\centering
\includegraphics[width=0.95\textwidth]{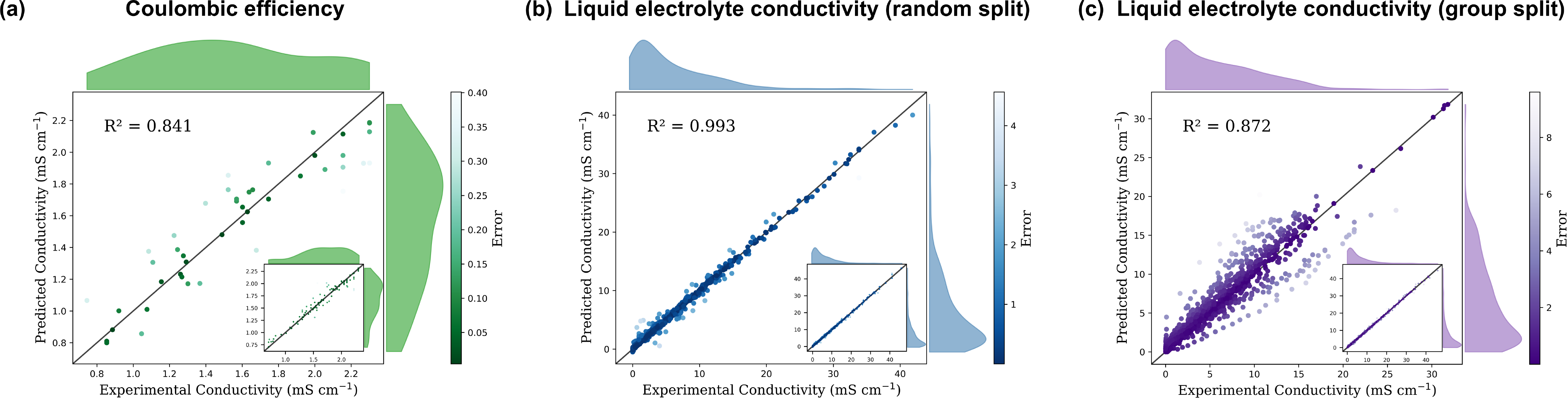}
\caption{\textbf{Regression plots for electrolyte formulation property prediction using Uni-ELF.} \textbf{(a)} Results of the Coulombic efficiency dataset. \textbf{(b,c)} Liquid electrolyte conductivity dataset, with \textbf{(b)} representing the random split and \textbf{(c)} the group split. The regression plots show the parity between experimental and predicted values in the test sets, with insets showing the results in the training sets. To illustrate data distribution, kernel density estimation is displayed at the top and right of each plot. The color gradients in the plots indicate the magnitude of prediction errors.}
\label{fig:reg}
\end{figure*}
\subsection{Formulation-Level Tasks}
To validate the efficacy of our multi-level representation learning architecture and transfer learning strategy for predicting solution structures, we applied the framework to the prediction of electrolyte formulation properties. Specifically, we reviewed and corrected two datasets from original sources: one on Coulombic efficiency (CE) for lithium metal anode batteries\cite{cuiyi_panas_2023}, and another on electrolyte conductivity\cite{bradford_chemistry-informed_2023}. For the Coulombic efficiency dataset, we removed an entry with a repeated ratio but different measurement methods and values, and corrected errors in some ratios and molecular information. This resulted in a refined dataset consisting of 149 entries of logarithmic Coulombic efficiency (LCE, defined as \(-\log(1-\text{CE})\)). For the conductivity dataset, errors were similarly corrected, and polymers were filtered out to focus on liquid electrolytes. The final conductivity dataset, curated at various temperatures, consisted of 2,588 entries.

Both datasets were split into training and test sets using a 7:3 ratio. Additionally, to evaluate the model’s ability to predict novel formulation systems, we employed an additional group split method for the conductivity dataset. In this method, data from formulation systems containing identical sets of molecular species were grouped and then randomly divided into training and test sets according to these groups. We utilized five-fold cross-validation during training to enhance the model’s robustness. The final model was an ensemble of the five models trained in each fold, with performance metrics derived from the averaged test set predictions.

We establish several baseline methods for constructing formulation fingerprints at both the molecular and formulation levels, utilizing XGBoost \cite{Chen_2016} for regression prediction. These methods include: one-hot encoding for all types of molecules in the dataset, where the formulation fingerprint contains only molecular species and ratio information without any molecular or solution structure details; Morgan fingerprints for encoding molecular structures \cite{rogers2010extended}; and Uni-Mol fingerprints derived from the Uni-Mol pre-trained model \cite{zhou2023unimol}, which do not dynamically adjust features. To enhance predictive accuracy in the electrolyte scenario, we partition the formulation fingerprint into solvent and salt components. Specifically, the fingerprints of molecules or ions are weighted by their molar ratios to generate the corresponding parts’ fingerprints, which are then concatenated to form the complete formulation fingerprint. Additionally, for the conductivity dataset, temperature is incorporated as a one-dimensional feature within the formulation fingerprint.

A summary of the performance of various molecular representation schemes on different tasks is provided in Table \ref{tab:rmse_results}. Notably, all the discussed schemes significantly outperform recent work by Kim et al. \cite{cuiyi_panas_2023}. Across all tasks, a consistent trend in performance is observed: the pre-trained Uni-ELF model achieves the best results, followed by the non-pre-trained Uni-ELF model, then the Uni-Mol fingerprint, Morgan fingerprint, and finally, the one-hot embedding. For instance, on the LCE dataset, the pre-trained Uni-ELF model achieves an RMSE of 0.184, reducing the error by approximately 14\% compared to the non-pre-trained Uni-ELF model, which has an RMSE of 0.215. Similarly, for the conductivity dataset, the pre-trained Uni-ELF model achieves an RMSE of 0.50 mS/cm (random split) and 2.15 mS/cm (group split), reducing the error by about 6\% and 13\%, respectively, compared to the non-pre-trained Uni-ELF model.

The alignment of these performance results with intuitive expectations is evident. One-hot embeddings, being simple numerical representations without structural information, perform the worst. Morgan fingerprints, which capture some molecular-level features, show moderate improvement. Uni-Mol fingerprints, containing richer molecular structures, further enhance performance. The superior outcomes of the non-pre-trained Uni-ELF model over Uni-Mol fingerprints with XGBoost highlight the efficacy of the transformer-based Uni-ELF architecture. Finally, the pre-trained Uni-ELF model, which incorporates even richer formulation-level structural information, achieves the best performance across all tasks.

As illustrated in Figure \ref{fig:reg}, the agreement between Uni-ELF predictions and experimental results is evident. Specifically, Figure \ref{fig:reg}(c) shows that while a group split may introduce more deviations—since some tested data belong to groups not present in the training data—the predictions still maintain a consistent trend. This demonstrates the robustness of the Uni-ELF model in handling diverse datasets and its ability to generalize well even under challenging conditions.

In conclusion, the pre-trained Uni-ELF model sets a new benchmark for predictive accuracy in this domain, demonstrating the critical importance of capturing comprehensive molecular and formulation-level information for superior performance in downstream tasks.

\section{Applications}
\begin{figure*}[h!]
\centering
\includegraphics[width=0.99\textwidth]{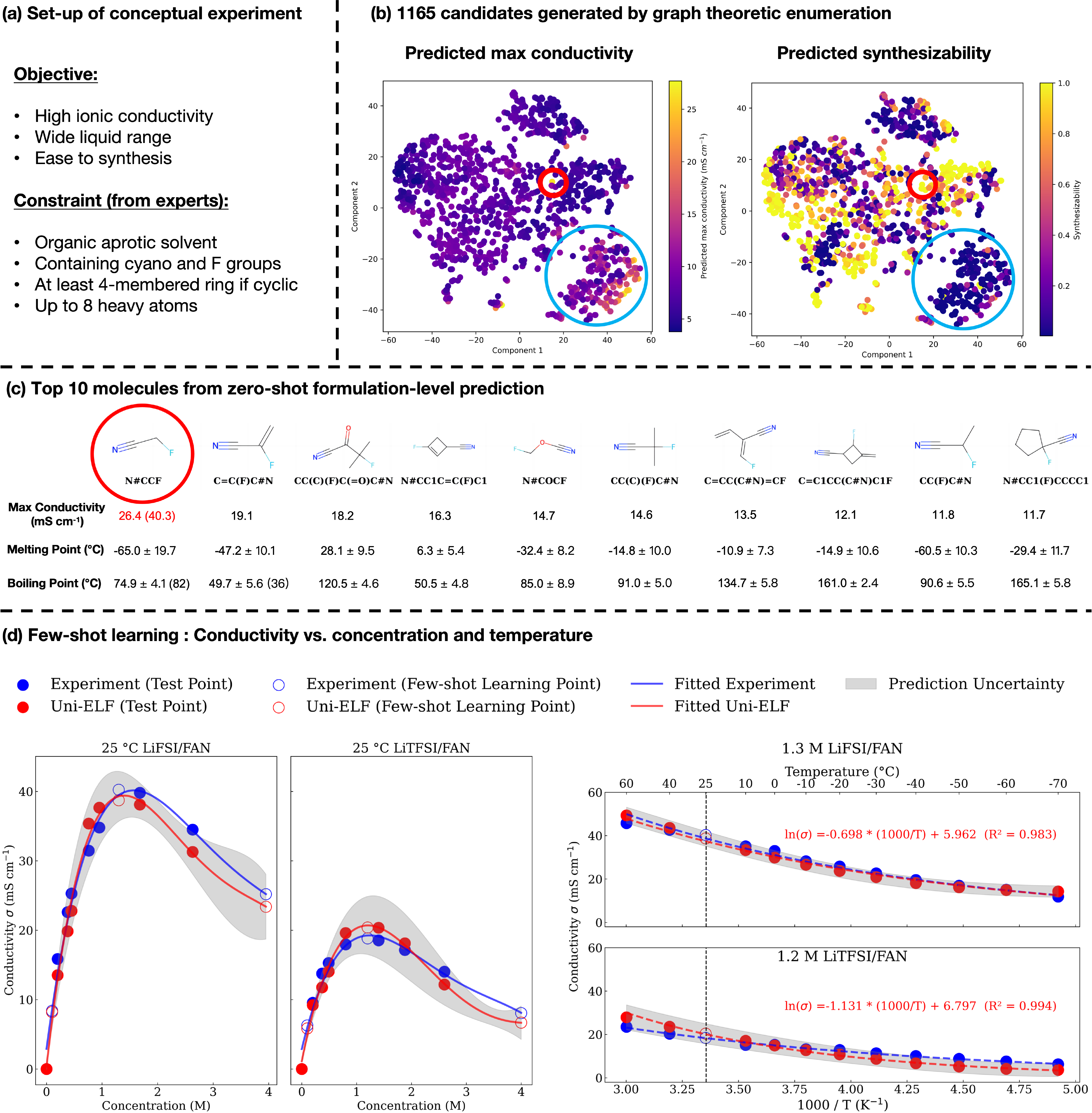}
\caption{\textbf{Conceptual electrolyte design using Uni-ELF.}
\textbf{a,} Set-up of the conceptual experiment: The objective is to achieve high ionic conductivity, a wide liquid range, and ease of synthesis. The molecular space to search is constrained by some practical expert criteria.
\textbf{b,} 1,165 candidates generated by graph-theoretic enumeration, visualized using t-SNE\cite{van2008visualizing} to reduce molecular representations to two dimensions, and color-coded by predicted maximum conductivity and synthesizability. The red circle highlights the high-conductivity FAN (fluoroacetonitrile) molecule discovered by the model, while the blue circle highlights a series of four-membered ring molecules with high predicted conductivity but low predicted synthesizability, which were thus screened out.
\textbf{c,} Top 10 molecules from zero-shot formulation-level prediction, emphasizing FAN's superior performance. Positive and negative values indicate model-predicted standard deviations, with parentheses showing experimental values.
\textbf{d,} Few-shot learning: Conductivity vs. concentration and temperature for LiFSI/FAN and LiTFSI/FAN systems. The model accurately predicts the conductivity-concentration relationship using data from only three experimental points: the initial concentration point (0.1 M), the final concentration point (4 M), and the peak conductivity concentration point (1.3 M for LiFSI/FAN and 1.2 M for LiTFSI/FAN) predicted by the model, which notably aligns with the experimental results. For the conductivity-temperature relationship, the model accurately predicts the high conductivity performance of FAN at low temperatures, fitting well to the Arrhenius relationship (red text).}
\label{fig:fan}
\end{figure*}

Although a comprehensive computational-experimental validation is deferred to future studies due to associated costs, we present the potential of Uni-ELF for molecular and formulation design via a conceptual application. In this context, we illustrate the rediscovery of fluoroacetonitrile (FAN), a high-conductivity solvent system recently reported by Lu et al. in Nature \cite{nature2024-fanxl}, with minimal constraints on the molecular search space. We start by constraining the search space to molecules that are organic aprotic solvents containing cyano and fluoro groups, incorporating at least a four-membered ring if cyclic, and comprising up to eight heavy atoms. Compatibility with high-voltage cathodes is ensured by the inclusion of electron-withdrawing cyano (\ce{-C#N}) groups, while anode compatibility is facilitated by fluorinated (\ce{-F}) groups. 

As shown in Figure \ref{fig:fan}(a), the conceptual experiment applies three objectives: high ionic conductivity for fast charging, wide liquid range for the solvent, and ease of synthesis, alongside four constraints mentioned above. To address these constraints, we begin by employing graph theory to enumerate potential molecules, starting with formonitrile (\ce{H-C#N}). Utilizing a breadth-first search (BFS), we progressively add carbon (C), oxygen (O), or fluorine (F) atoms to the chain. Duplicates are filtered out on the basis of graph isomorphism, and the search continues until we generate chain molecules containing up to eight heavy atoms. Subsequently, we enumerate all possible configurations to form 4-6 membered rings, again eliminating duplicates. Unstable structures, such as those containing O-O bonds, or proton groups unsuitable for use as electrolytes, such as carboxyl groups, are discarded, resulting in a set of 1,165 candidate molecules. This entire search process is completed in under 30 seconds on a standard computer CPU.

Following the identification of 1,165 candidate molecules, we employed Uni-ELF models to efficiently screen these compounds for their molecular and formulation properties. Molecular level properties, including melting point, boiling point, synthetic accessibility (synthesizability), and electrolyte conductivity, are predicted using Uni-ELF trained by publicly available data, with little direct information of the 1,165 candidates. At the formulation level, we generated a grid of 120 formulation points by systematically combining each molecule with LiPF$_6$, LiTFSI, and LiFSI salts at varying concentrations. This approach enabled us to predict the room temperature conductivity for each formulation, providing a robust basis for chemists to establish evaluation criteria for subsequent experimental testing.

We filter and rank the molecules using the following criteria: (1) predicted melting point $\leq 40^\circ$C and boiling point $\geq 40^\circ$C; (2) presence in PubChem\cite{kim2023pubchem} or CAS, or synthetic accessibility probability $\geq 90\%$; (3) highest predicted room temperature conductivity among all formulations. Through prediction and screening, we identify the top 10 candidate molecules, shown in Figure \ref{fig:fan}(c). The top-ranked molecule is fluoroacetonitrile (FAN), with a predicted maximum room temperature conductivity of 26.35 mS/cm, significantly surpassing the second-ranked molecule. According to Lu et al.\cite{nature2024-fanxl}, FAN exhibits a high ionic conductivity of 40.3 mS/cm as a lithium-ion electrolyte. Notably, FAN was not present in the model's training datasets, suggesting that the model independently identified FAN as a promising electrolyte material.

Using additional data published by Lu et al.\cite{nature2024-fanxl}, we further explore the few-shot generalization capability of Uni-ELF. As shown in Figure \ref{fig:fan}, after few-shot learning with 3 data points—corresponding to the endpoint concentrations (0.1 and 4 M) and the concentration with the highest predicted conductivity—the model accurately predicts the conductivity-concentration relationships for both LiFSI/FAN and LiTFSI/FAN systems. 
We perform 10-fold cross-validation, train 10 models, and average their predictions. The uncertainty values are provided by the standard deviation, and the curves fitting the experimental and predicted values are displayed using a fourth-order polynomial. 
In the low concentration region, the two curves almost coincide; in the high concentration region, the model's prediction uncertainty is higher, which can be attributed to fewer high concentration data points in the training dataset.

Furthermore, we used the model to predict the conductivity-temperature relationship at the concentration with the highest conductivity for both LiFSI/FAN and LiTFSI/FAN systems, as shown in Figure \ref{fig:fan}. Using only 1 data point—the room temperature data—for retraining, the model successfully predicts the high ionic conductivity performance of the LiFSI/FAN system at low temperatures. We fit the model predictions for both systems using the Arrhenius relationship, obtaining $\ln(\sigma) = -0.698 \times \left(\frac{1000}{T}\right) + 5.962$ ($R^2 = 0.983$) for LiFSI/FAN and $\ln(\sigma) = -1.131 \times \left(\frac{1000}{T}\right) + 6.797$ ($R^2 = 0.994$) for LiTFSI/FAN. This indicates that the model effectively learns the Arrhenius relationship of conductivity with temperature from the original dataset and successfully transfers this knowledge to the FAN system. 

In summary, Uni-ELF demonstrates the ability to effectively integrate physical modeling and publicly available experimental data to achieve accurate predictions of molecular and formulation-level properties. This predictive accuracy enabled the model to independently rediscover high-performing molecules like FAN, underscoring Uni-ELF's potential in advancing electrolyte design. Furthermore, the capability of Uni-ELF to perform few-shot learning suggests its potential for low-cost, efficient optimization of high-dimensional formulation spaces, presenting a promising avenue for future integration into robotic automated experimentation.

\section{Conclusion and Outlook}
In this work, we have introduced Uni-ELF, a multi-level representation learning framework designed to advance the formulation and optimization of electrolytes for lithium batteries. By leveraging a two-stage pretraining approach—reconstructing three-dimensional molecular structures using the Uni-Mol model and predicting statistical structural properties from molecular dynamics simulations—we have demonstrated significant improvements in predictive capabilities for both molecular and formulation properties.

Our results show that Uni-ELF outperforms current state-of-the-art methods in predicting key properties such as melting point, boiling point, synthesizability, conductivity, and Coulombic efficiency. Notably, Uni-ELF can be seamlessly integrated into an automatic experimental design workflow, bridging the gap between computational predictions and experimental validation.

Looking forward, the methodology presented here holds promise for broader applications beyond electrolyte design. For example, this approach could be extended to other areas requiring formulation-level prediction or generation, such as the design of pharmaceuticals and the extraction of formulation information from spectral data. We are optimistic that further refinement and validation of this framework will enhance its utility and impact across various scientific and engineering domains.

\section*{Data and Code Availability}
Data and code used in this work will be made publicly available after the paper is published. For trial use of Uni-ELF, please refer to the Bohrium App at \url{https://bohrium.dp.tech/apps/uni-elf}.

\section*{Conflict of Interests}
DP Technology holds intellectual property rights pertinent to the research presented herein.


\newpage
\appendix

\onecolumn

\newpage
\section*{Supplementary Information}
\subsection*{Formulation-Level Model Architecture}
The Uni-ELF backbone consists of three main parts: the temperature embedding block, the formulation transformer encoder, and the molecular pair RDF block (see Figure~\ref{fig:model}(c)).

\textbf{Temperature Embedding Block:}
The temperature input is transformed using a Gaussian kernel followed by layer normalization \cite{ba2016layer}. The Gaussian kernel embeds the temperature value into a 256-dimensional feature utilizing 512 Gaussian basis functions, with a nonlinear projection reducing the dimensionality from 512 to 256. The means and standard deviations of these Gaussian basis functions are initialized uniformly between -80 and 150, and 0 and 230, respectively. Subsequently, layer normalization stabilizes the temperature embeddings' means and variances, ensuring a stable distribution for further processing.

\textbf{Formulation Transformer Encoder:}
The formulation transformer encoder integrates molecular representations with their corresponding normalized molar ratios. Initially, the molecular representations are scaled by their normalized molar ratios. These scaled representations are then fed into a multi-head attention layer \cite{vaswani2017attention}, which employs 64 attention heads to compute interaction scores between the input features. Each attention head has a dimension of 8, resulting in a total embedding dimension of 512. The output from the attention mechanism is then normalized using layer normalization and combined with the input of the attention layer via a residual connection \cite{he2016deep}. Following this, a feedforward network comprising two linear layers is applied. The first linear layer expands the input dimension from 512 to 2048. The output of this layer is activated using a GELU function \cite{hendrycks2016gaussian} and then mapped back to 512 dimensions by the second linear layer. The feedforward network's output undergoes another layer normalization and is added to its input through an additional residual connection. After three such layers, the refined molecular representations are aggregated by weighted summation and concatenated with the temperature embeddings to form the final formulation representation.

\textbf{Molecular Pair RDF Block:}
During pretraining, the molecular pair RDF block refines the formulation representation by incorporating radial distance information. The radial distance is encoded using another Gaussian kernel, which transforms the distance into 128 Gaussian basis functions with means and standard deviations uniformly initialized between 0 and 1.5. A nonlinear layer further reduces the 128-dimensional encoding to a 64-dimensional radial distance embedding, followed by layer normalization. The molecular pair representation, which has 64 dimensions corresponding to the number of attention heads, is then concatenated with the radial distance embeddings. This combined representation is fed into a linear layer with an input dimension of 128 (64 from the molecular pair representation and 64 from the radial distance embeddings) and an output dimension of 128. The output of this linear layer is activated using a tanh function \cite{lecun2015deep} to introduce non-linearity. Finally, a second linear layer maps these 128-dimensional features to a single scalar value, representing the RDF prediction for a molecular pair.

\subsection*{Formulation-Level Pretraining}
For the electrolyte systems used in pretraining, we select classic binary mixtures of linear carbonate and cyclic carbonate solvents. The linear carbonates include ethylmethyl carbonate (EMC), dimethyl carbonate (DMC), and diethyl carbonate (DEC), while the cyclic carbonates consist of ethylene carbonate (EC) and propylene carbonate (PC). For each type of solvent component, one molecule is selected, and five points are uniformly generated within the molar fraction range of 0-1. At each grid point, random perturbations are applied within a molar fraction range of 0.15. The lithium salts used are either lithium hexafluorophosphate (LiPF\(_6\)) or lithium bis(fluorosulfonyl)imide (LiFSI), with salt molality chosen as 0.5, 1.0, and 1.5 mol/kg solvent, resulting in 180 formulations for classical molecular dynamics simulations.

Each formulation contains up to four types of molecules or ions (with the salt split into cations and anions), resulting in up to ten pairs of molecular RDFs. Each pair of molecular RDFs includes approximately 900 data points (r, g(r)) within the range of radial distance r from 0 to 2.0 nm, ultimately generating a dataset of approximately 160,000 data points. For the purpose of learning the inter-molecular interactions in the pretraining procedure, all inner-molecular contributions of RDFs are ignored. The dataset is split into training, validation, and test sets in a ratio of 8:1:1 for pretraining. The model is trained to minimize the RMSE between the predicted and true RDFs.

\subsection*{Details of Molecular Dynamics Simulations}
The Molecular Dynamics (MD) simulations of electrolyte formulations were carried out using GROMACS\cite{Abraham2015GROMACSHP} package. Parameters of Generated Amber force field (GAFF) for all electrolyte solutes and solvents were obtained using Antechamber\cite{WANG2006247} in Ambertools23 package and ACPYPE software\cite{SousaDaSilva2012}. Atomic partial charges were generated via RESP scheme as follows: All molecules are optimized in B3LYP/6-311g(d,p) DFT level using Gaussian 16 software\cite{g16}, where solvent effect is introduced using PCM method. Then, RESP charges are fitted from the optimized geometry and wave function using Multiwfn software\cite{https://doi.org/10.1002/jcc.22885}.

All electrolyte molecules were put intis used for possible pressure control. The particle-mesh Ewald (PME) method is used for electrostatics. Atoms linking with hydrogen atoms are restrained by LINCS algorithm.

The systems are firstly equilibrated at 298 K, 1000 atm in NPT ensemble for 200 ps with a time step of 2 fs to reach a reasonable density. A further pre-equilibrium process of 95000 ps in total is conducted, which is consisted of several simulated annealing processses and NPT processes. After the pre-equilibrium process, the systems are adjusted to the average density for the generation of a 5000 ps trajectory files in NVT ensemble. The details of simulation settings are listed in Table 1.

\begin{table}[ht]
\centering
\caption{Main simulation settings of molecular dynamics}
\label{tab:simulation_settings}
\resizebox{\textwidth}{!}{
\begin{tabular}{clccccc}
\hline
\textbf{Step number} & \textbf{Description} & \textbf{Ensemble} & \textbf{Temperature (K)} & \textbf{Pressure (atm)} & \textbf{Time step (fs)} & \textbf{Simulation steps} \\ \hline
1 & Energy minimization & -- & -- & -- & -- & $< 50000$ \\
2 & Equilibrium & NVT & 298 & -- & 1 & 5000 \\
3 &  & NPT & 298 & 1000 & 2 & 100000 \\
4 & Anneal & NVT & 298-363-298 & -- & 2 & 1000000 \\
5 & Equilibrium & NPT & 298 & 1 & 2 & 500000 \\
6 & Anneal & NVT & 298-363-298 & -- & 2 & 2500000 \\
7 & Equilibrium & NPT & 298 & 1 & 2 & 500000 \\
8 & Scaling configuration to average density & -- & -- & -- & -- & -- \\
9 & Trajectory generation & NVT & 298 & -- & 2 & 2500000 \\ \hline
\end{tabular}
}
\end{table}
The radial distribution functions (RDFs) were calculated using GROMACS, ranging from 0 to the radius of the simulation box with a bin width of 0.002 nm. For components A and B, the RDF between them is computed using the following equation:

\[
g_{AB}(r) = \frac{\langle \rho_B(r) \rangle}{\langle \rho_B \rangle_{local}} = \frac{1}{\langle \rho_B \rangle_{local}} \frac{1}{N_A} \sum_{i \in A}^{N_A} \sum_{j \in B}^{N_B} \frac{\delta(r_{ij} - r)}{4\pi r^2}
\]

where \( g_{AB}(r) \) is the radial distribution function between components A and B at a distance \( r \), \( \langle \rho_B(r) \rangle \) is the average local density of component B at distance \( r \) from a particle of component A, \( \langle \rho_B \rangle_{local} \) is the particle density of type B averaged over all spheres around particles A with radius \( r_{max} \), \( N_A \) is the number of particles of component A, \( N_B \) is the number of particles of component B, \( r_{ij} \) is the distance between particle \( i \) of component A and particle \( j \) of component B, and \( \delta(r_{ij} - r) \) is the Dirac delta function, which is 1 when \( r_{ij} = r \) and 0 otherwise.

\subsection*{Details of the Downstream Tasks}
The experimental evaluation methods vary among the different approaches we compare. To ensure fair comparisons, we align our evaluation settings as closely as possible with those of the compared methods, particularly in the way training and test sets are divided. If the original method provides specific training and test sets, we use those. If only the division ratio is stated, we split the data using three random seeds and report the average metrics of three test results. During training, we employ five-fold cross-validation to enhance the robustness of the model.  In each fold, the model undergoes training for 200 epochs, selecting the checkpoint demonstrating optimal performance on the validation set. The final model combines the predictions from the five models trained in each fold, averaging these predictions to determine the performance metrics. The comparison results are listed in Table \ref{single-performance}.

\begin{table}[h]
\label{single-performance}
\caption{Prediction performances of various molecular properties}
\centering
\resizebox{\textwidth}{!}{
\begin{tabular}{lccccccccccccccccccccccc}
\toprule
Dataset & \multicolumn{2}{c}{Melting Point} & \multicolumn{2}{c}{Boiling Point} & \multicolumn{2}{c}{Vapor Pressure} & \multicolumn{2}{c}{Dielectric Constant} & \multicolumn{2}{c}{Refractive Index} & \multicolumn{2}{c}{Density} & Synthesizability\\

\cmidrule(lr){2-3} \cmidrule(lr){4-5} \cmidrule(lr){6-7} \cmidrule(lr){8-9} \cmidrule(lr){10-11} \cmidrule(lr){12-13} \cmidrule(lr){14-14} \\

Dataset Size & \multicolumn{2}{c}{19572} & \multicolumn{2}{c}{5435} & \multicolumn{2}{c}{2713} & \multicolumn{2}{c}{1220} & \multicolumn{2}{c}{7243} & \multicolumn{2}{c}{8905}  & 126405 \\
Training Set Ratio & \multicolumn{2}{c}{0.90} & \multicolumn{2}{c}{0.75} & \multicolumn{2}{c}{0.75} & \multicolumn{2}{c}{0.70} & \multicolumn{2}{c}{0.50} & \multicolumn{2}{c}{0.53} & 0.79\\ 
Reference Work & \multicolumn{2}{c}{Mi et al.\cite{mi_melting_2021}} & \multicolumn{2}{c}{Mansouri et al.\cite{mansouri_opera_2018}} & \multicolumn{2}{c}{Mansouri et al.\cite{mansouri_opera_2018}} &\multicolumn{2}{c}{Bouteloup et al.\cite{bouteloup_predicting_2019}} &\multicolumn{2}{c}{Bouteloup et al. \cite{bouteloup_improved_2018}} &\multicolumn{2}{c}{Mathieu et al. \cite{mathieu_reliable_2016}} & \multicolumn{1}{c}{Lee et al. \cite{lee_emin_2024}} \\

Scores & RMSE & R$^2$ & RMSE & R$^2$ & RMSE & R$^2$ & RMSE & R$^2$ & RMSE & R$^2$ & RMSE & R$^2$ & AUC \\
\midrule
Ref     & 36.88 & 0.830 & 22.08 & 0.93  & 1.00 & 0.92  & 5.03 & 0.91  & 0.0136 & 0.950 & 0.028 & 0.989  & 0.955 \\ 
Uni-ELF & \textbf{34.31} & \textbf{0.857} & \textbf{13.49} & \textbf{0.975} & \textbf{0.79} & \textbf{0.951} & \textbf{2.70} & \textbf{0.966} & \textbf{0.0082} & \textbf{0.982} & \textbf{0.025} & \textbf{0.992}  & \textbf{0.965} \\

\bottomrule
\end{tabular} 
}

\end{table}
\paragraph{Melting Point}
Predicting the melting point has long been a challenging task in cheminformatics\cite{tetko_how_2014}. The highest quality dataset available, to the best of our knowledge, is the 2014 Jean-Claude Bradley Open Melting Point Dataset\cite{Bradley2014}, which comprises 19,933 entries and 28,645 measurement records, some of which are marked as erroneous.

The state-of-the-art method employed natural language processing (NLP) techniques to process molecular SMILES strings \cite{mi_melting_2021}. Allocating 10\% of the dataset for testing, the model achieved a R² of 0.830 and a root mean square error (RMSE) of 36.88 °C. Our dataset preprocessing approach closely aligns with this method but includes several refinements: erroneous records are excluded, inorganic compounds without carbon are omitted, and entries with duplicate measurements deviating from the mean by more than five degrees are removed. The remaining entries are averaged into a single record, resulting in a refined dataset of 19,572 unique records.

\paragraph{Boiling Point and Vapor Pressure}
We use the database for boiling points and vapor pressures from the OPERA model\cite{mansouri_opera_2018}, which employed a QSPR modeling approach. OPERA designated 75\% of the data for training, achieving a test R² of 0.93 and an RMSE of 22.08 °C across 5,435 records with boiling points measured at 760 mm Hg. For the vapor pressure dataset, which includes 2,713 records, the test R² reached 0.92 with an RMSE of 1.00 Log mm/Hg. Our method substantially outperforms the OPERA model on these metrics, achieving a test R² of 0.975 with an RMSE of 13.49 °C for boiling points, and a test R² of 0.951 with an RMSE of 0.79 Log mm/Hg for vapor pressures.

\paragraph{Dielectric Constant, Refractive Index, and Density}
Bouteloup and Mathieu\cite{bouteloup_predicting_2019,bouteloup_improved_2018,mathieu_reliable_2016} developed a series of QSPR models based on physical equations for predicting dielectric constant, refractive index, and density properties. Using the same training-test split as their studies, our method demonstrates superior performance on the test set.. For the dielectric constant, 
we achieve an R² of 0.966 and an RMSE of 2.70; for the refractive index, an R² of 0.982 and an RMSE of 0.082; and for the density, an R² of 0.992 and an RMSE of 0.025.

\paragraph{Synthesizability}
For the synthesizability dataset, Lee et al.\cite{lee_emin_2024} randomly selected 100,000 entries as a training set and utilized chemical descriptors as molecular features to test classification accuracy on the remaining data, achieving a peak Area Under the Curve (AUC) of 0.955. Using the same evaluation settings, our approach reaches an AUC of 0.965 on the test set, once again surpassing methods that rely on specific feature engineering.

\paragraph{Baseline Methods on Formulation-Level Tasks}
For all baseline methods using XGBoost, we performed hyperparameter optimization using Optuna\cite{akiba2019optuna} on the validation sets in a five-fold cross-validation. The hyperparameters optimized included the number of estimators (100 to 1500), maximum depth (3 to 9), learning rate (0.0001 to 0.3), column sampling by tree (0.1 to 1.0), and alpha (1 to 10). The optimization aimed to minimize the root mean squared error (RMSE) by tuning these parameters.

\paragraph{Additional Benchmarks}
We also benchmark Uni-ELF against leading deep learning-based models in five property prediction tasks (dielectric constant, density, melting point, boiling point, and refractive index). Uni-ELF demonstrates the best performance in four out of the five tasks (Table \ref{bechemark}). To maintain consistency, a 9:1 training-to-test set ratio is used across all comparisons. For each model, a comprehensive grid search over hyperparameters—batch size, learning rate, and embedding dimensions—is performed using a 3x6x3 grid. The final results are reported using the best-performing hyperparameter set identified through this search, ensuring that each model is evaluated under optimal performance conditions.

\begin{table}[ht]
\label{bechemark}
\caption{Performance of the Uni-ELF and other leading deep learning-based models in dielectric, density, melting point, boiling point, and refractive index datasets, with the best RMSE and R$^2$ scores denoted in bold and second best underlined.}
\centering
\begin{tabular}{lccccccccccc}
\toprule
Datasets & \multicolumn{2}{c}{Dielectric Constant}  & \multicolumn{2}{c}{Density} & \multicolumn{2}{c}{Melting Point} & \multicolumn{2}{c}{Boiling Point} & \multicolumn{2}{c}{Refractive Index}\\
\cmidrule(lr){2-3} \cmidrule(lr){4-5} \cmidrule(lr){6-7} \cmidrule(lr){8-9} \cmidrule(lr){10-11}\\
Scores & RMSE & R$^2$ & RMSE & R$^2$ & RMSE & R$^2$ & RMSE & R$^2$ & RMSE & R$^2$\\
\midrule
XGBoost\cite{Chen_2016} & $8.737$ & $0.685$ & $0.0915$ & $0.897$& $50.343$ & $0.693$ & $47.710$ & $0.731$ & $0.0219$ & $0.886$\\
D-MPNN\cite{dmpnn} & $4.081$ & $0.932$ & $\textbf{0.0229}$ & $\textbf{0.994}$ & $44.205$ & $0.774$ & $24.554$ & $0.931$ & $0.0110$ & $0.972$\\
GAT\cite{gat} & $7.018$ & $0.832$ & $0.0964$ & $0.894$& $48.087$ & $0.738$ & $19.177$ & $0.956$ & $0.0183$ & $0.935$\\
GCN\cite{gcn} & $5.214$ & $0.898$ & $0.0796$ & $0.918$& $42.358$ & $0.785$ & $18.713$ & $0.959$ & $0.0182$ & $0.944$\\
PAGTN\cite{pagtn} & $5.216$ & $0.904$ & $0.0405$ & $0.987$& $42.292$ & $0.791$ & $17.219$ & $0.968$ & $0.0236$ & $0.920$\\
Uni-ELF & $\mathbf{3.219}$ & $\mathbf{0.953}$ & $\underline{0.0257}$ & $\underline{0.992}$ & $\mathbf{34.312}$ & $\mathbf{0.857}$ & $\mathbf{15.252}$ & $\mathbf{0.971}$ & $\mathbf{0.0087}$ & $\mathbf{0.982}$\\
\bottomrule
\end{tabular}
\end{table}


\end{document}